\documentclass[sn-mathphys,Numbered]{sn-jnl}

\usepackage{graphicx}%
\usepackage{multirow}%
\usepackage{amsmath,amssymb,amsfonts}%
\usepackage{amsthm}%
\usepackage{mathrsfs}%
\usepackage[title]{appendix}%
\usepackage{xcolor}%
\usepackage{textcomp}%
\usepackage{manyfoot}%
\usepackage{booktabs}%
\usepackage{algorithm}%
\usepackage{algorithmicx}%
\usepackage{algpseudocode}%
\usepackage{listings}%
\usepackage{natbib}

\theoremstyle{thmstyleone}%
\theoremstyle{thmstyletwo}%

\theoremstyle{thmstylethree}%
\begin{document}

\title[Article Title]{Characterization of the phonon sensor of the CRYOSEL detector with IR photons}


\author*[1]{ \sur{H. Lattaud}}\email{lattaud@ip2i.in2p3.fr}
\author[1]{ \sur{E. Guy}}
\author[1]{ \sur{J. Billard}}
\author[1]{ \sur{J. Colas}}
\author[1]{ \sur{M. De Jésus}}
\author[1]{ \sur{J. Gascon}}
\author[1]{ \sur{A. Juillard}}
\author[2]{\sur{S. Marnieros}}
\author[2]{\sur{C. Oriol}}

\affil*[1]{\orgdiv{Univ. Lyon, Univ. Lyon 1}, \orgname{CNRS/IN2P3, IP2I Lyon }, \orgaddress{\street{4 rue Enrico Fermi}, \city{Villeurbanne}, \postcode{F-69622}, \country{France}}}
\affil[2]{\orgdiv{Université Paris-Saclay}, \orgname{CNRS/IN2P3, IJCLab}, \orgaddress{\city{ Orsay}, \postcode{91405}, \country{France}}}


\abstract{The sensitivities of light Dark Matter (DM) particle searches with cryogenic detectors are mostly limited by large backgrounds of events that do not produce ionization signal. The CRYOSEL project develops a new technique where this background in a germanium cryogenic detector is rejected by using the signals from a Superconducting Single Electron Device (SSED) sensor designed to detect the phonons emitted through the Neganov-Trofimov-Luke effect by the e$^-$h$^+$ pairs as they drift in a close-by very high-field region. A tag on signals from this device should suppress the heat-only background. The measurement of the response to IR laser pulses of the first CRYOSEL prototype show the relevance of such sensor technology.}


\maketitle

\section{Introduction}\label{sec:intro}
The search for direct interactions of Dark Matter (DM) particles, such as Weakly Interacting Massive Particles (WIMPs), with the nuclei or the electronic cloud has been ongoing for decades. The absence of signal in the explored energy range $[\text{GeV.c}^2,\text{TeV.c}^2]$~\cite{rev,rev-bertone,xenon,xenon-lux,xenon-panda} has led to the development of new techniques to lower the energy threshold down to the eV range~\cite{lowmass,cresst,SuperCDMS-HVeV,CPD,damic,sensei,damic-m}. In this newly probed energy range, additional backgrounds that are not fully understood are emerging~\cite{excess}. Cryogenic experiments observe a low-energy excess of events that are not associated with ionization signals~\cite{nbsi-migdal}. Here, this background will be referred to as \textit{Heat-Only} (HO) events. Recent studies hint towards sudden stress releases in the detectors~\cite{cresst-crack} as a plausible origin of this background. In the former EDELWEISS experiment, the HO events have been the dominating background at low energy, limiting the sensitivity of past searches~\cite{edwiii,red30,nbsi-migdal}.

In order to mitigate the effects of the HO background, the member of the CRYOSEL project are developing a new type of sensor called the \textit{Superconducting Single Electron Device} (SSED), that can be added to its germanium (Ge) bolometers equipped with Ge-NTD (Neutron Transmutation Doped) thermistors. The SSED will be used to tag the athermal phonons emitted via the Neganov-Luke-Trofimov (NTL) effect~\cite{NTL} created by the drift of e-/h+ pairs. The electrodes covering the detector are designed as point-contact Ge to concentrate field lines in front of the SSED, in order to maximize its sensitivity to events with charge and minimize the probability to trigger on HO event, assumed to be distributed more uniformly in the detector volume. The SSED is made of a small, single NbSi line (Fig.~\ref{fig1} top left), with a critical temperature $T_c$ for its transition from the superconducting to the normal resistive state well above the temperature $T_{op}$ at which is operated the detector, in order to further reduce the probability of triggering on HO events.
While the SSED should provide a sensitive tag for the presence of charge, the precise measurement of the energy of the event will be provided by the NTD sensor.

This paper presents the results obtained with the first working CRYOSEL prototype allowing the simultaneous operation of a SSED and NTD sensor, in addition to a charge readout on the electrode to provide redundancy. The absolute calibration of these three channels was performed. We present the measurement of their response to IR laser pulses of known energy for biases ranging from 0 V to 60 V for different values of $T_{op}-T_C$. This paper is organized as follows: the experimental setup and its energy calibration is described in Sec.~\ref{sec1}, the SSED response to varying biases and operating temperature is described in  Sec.~\ref{sec2} and a conclusion is drawn in Sec.~\ref{sec3}.

\section{Experimental setup and energy calibration}\label{sec1}

\begin{figure}[!h]
\begin{center}
$\vcenter{\hbox{\includegraphics[width=0.39\linewidth]{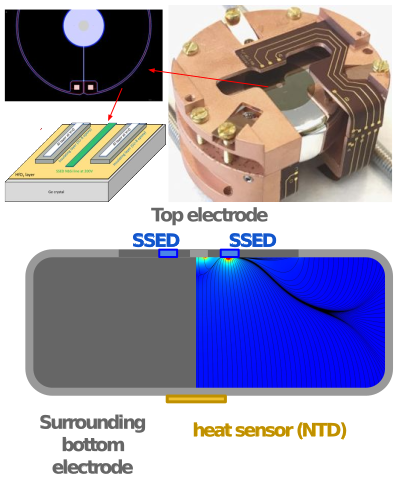}}}$
$\vcenter{\hbox{\includegraphics[width=0.60\linewidth]{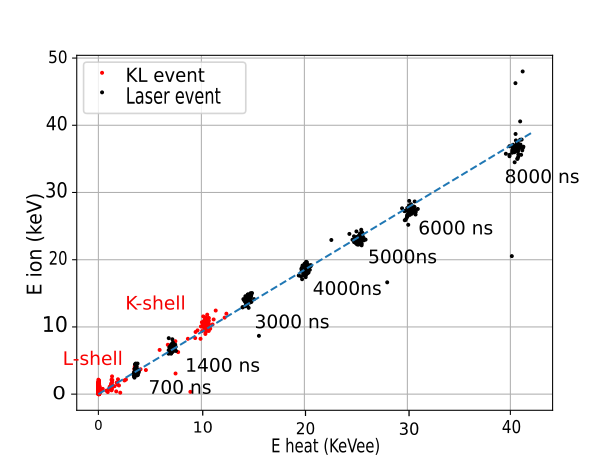}}}$

\caption{Top left: CRYO50 38~g detector equipped with a NbSi SSED  inside its copper holder, the surrounding electrode, the top electrode and the SSED are visible on the top side. A sketch of the SSED sensor highlight its geometry.
Bottom left: side  sketch view of the CRYO50 detector with the Ge crystal in dark gray, the surrounding and top Al electrode in dark gray, the SSED line is in blue. On one half of the sketch, the COMSOL MultiPhysics\textsuperscript\textregistered calculation~\cite{comsol} of the electric field intensity is shown. 
Right : the coincident ionization and heat energy measurement for the K and L lines (red) and laser pulses (black) at various pulse width, step from $700$ to $8000$ ns, the line shows the fitted linearity of the laser.}
\label{fig1}
\end{center}
\end{figure}

CRYO50 is a $10$ mm high and $30$ mm wide cylindrical high-purity Ge crystal with rounded edges, weighing $38$~g. A $30$ nm thick HfO$_2$ layer is deposited onto the crystal surfaces to prevent leakage current and to increase the sustained bias. A $150$ nm thick aluminum (Al) electrode is evaporated on the crystal, covering the majority of the surface except a 15mm wide Al-free disk  on top of the detector (see Fig~\ref{fig1} top left). A second Al electrode with two features, a $150$ nm thick and 1 mm wide plot and two line bracing the SSED sensor (described below), is evaporated in the center of the Al-free disk.

The SSED sensor has been lithographed  on this Al-free area, it consists of a $10$~$\mu$m thick Nb$_{x}$Si$_{(1-x)}$ line, shaped as a circle with a diameter of 5 mm with a pad at each ending, its superconducting transition temperature is  $T_c = 46$~mK. The electrodes and the SSED are electrically and thermally coupled to the thermal bath and to the electronic readout through an Al wire bonding. A future paper will explore with more details the fabrication process.

A  $2\times 2 \times 0.5$  mm$^2$ Ge-NTD heat sensor \cite{NTDpaper} is glued on an Al grid at the center of the bottom surface of the crystal and thermally coupled to the thermal bath and to the electronic.  Fig.~\ref{fig1} (left) shows a sketch of the detector with the Ge crystal in dark gray, the electrodes in light gray, the SSED in blue and the NTD in yellow, one of the halves shows the intensity lines of the electric field calculated using the COMSOL MultiPhysics\textsuperscript\textregistered software \cite{comsol} when grounding all the Al electrodes and biasing the SSED.

The signal readout consists of one SSED channel, one ionization channel from the enveloping electrode and one heat channel.
 As in Ref~\cite{red30}, the detector is operated in the Lyon dry dilution cryostat, a Hexadry-200 cryostat from Cryoconcept\textsuperscript\textregistered \cite{cryoconcept} and the cold and warm electronics are those described in Ref.~\cite{electroniclyon}. A $1650$~nm infrared laser diode from Aerodiode\textsuperscript\textregistered \cite{aerodiode} with a maximum optical power of $10$~mW and operated outside the cryostat is used to illuminate the detector with IR photons~\cite{selendis}. The average penetration length of such photon is $25$~cm ~\cite{absGe1,absGe2} ensuring that, given the multiple reflection on the copper detector housing and the aluminum electrodes, the crystal bulk is lighted.

The laser beam is guided through optical fibers thermalized at each stage of the cryostat. The end of the fiber is directly mounted on the detector's copper housing. The optical power is controlled using a set of attenuators. The laser was operated in pulsed mode. The energy deposited in each pulse was controlled via the pulse duration. The data was processed using the algorithm described in Ref.~\cite{MPS}. The information of the time of each laser pulse was recorded  using the trigger signal provided by the laser control board.

The detectors were activated for 16 hours using a strong AmBe neutron source before being mounted in the cryostat. This activation creates $^{71}$Ge which decays by electron capture with a half life of 11 days. The two X-rays lines observed in Fig.~\ref{fig1} (right, red points) at 1.3 keV and 10.37 keV correspond to decays to the L- and K-shells.  These peaks are used to provide an absolute calibration of the ionization and heat (NTD) channels. 

IR photons are injected in the crystal at a frequency of $f=0.2$~Hz.
The total phonon energy in the crystal is given by the expression: 
\begin{equation}\label{eq1}    
    E_{phonon} = E_{Laser}  + E_{Luke} = h\nu \times (N_0 + N_1)  + |\Delta V| \times N_{1}    
\end{equation}

$E_{Laser}$ is the total energy deposited by photons via the photoelectric effect and/or the creation of excitons and thermal phonons. Given the energy $h\nu=0.75$~eV of the IR photons here, each one can either create zero or one electron-hole pair, with $N_0$ and $N_1$ denoting the respective average number of such cases per each laser burst. When all charges have recombined either at the electrodes or in the crystal, their energy is converted back in phonons and contribute to the measured $E_{phonon}$. $E_{Luke}$ is an additional contribution to the total phonon energy coming from the drifting of the charges under the effect of a bias $|\Delta V|$. This effect is called the Neganov-Trofimov-Luke (NTL) effect \cite{NTL}. The equivalent expression for the signal from the L and K electron recoil signals used to calibrate the NTD sensor is $E_{phonon} = E_{recoil}(1+V/\epsilon_\gamma)$, with $\epsilon_\gamma=3$~eV for Germanium. The energy scale deduced from these electron recoil signal is, thus  $E_{heat} = E_{phonon}/(1+V/\epsilon_\gamma)$, expressed in keV-equivalent-electron ($\text{keV}_{\text{ee}}$)~\cite{red30}. This energy scale, chosen for this work, is the one that will be typically used in search for DM interactions with electrons, and can be compared directly to the calibrated signals from the charge readout (as in Fig.~\ref{fig1}).

the coincident ionization and heat measurement shown in Fig.~\ref{fig1} (right) is made with data recorded with the detector biased at $60$~V, the observed baseline heat (ionization) energy resolution is $153$ eV ($435$~eV) (RMS). The black points in Fig~\ref{fig1} (right) are coincident with laser pulses of different lengths. The charge and heat signals both increase linearly with the pulse width. As in Ref~\cite{selendis}, these data are used to determine the absolute number of charge and total energy deposited as a function of the selected laser pulse width.

\section{Results}\label{sec2}
The SSED is meant to reject HO events by tagging the presence of charges produced within the crystal. In this section, we present results on the response of the first operational SSED-equipped CRYOSEL detector to charges created by IR photons pulses. 
All but two channels, the laser and the SSED, were disregarded for this analysis and only events coming from the IR photons are considered by triggering on the laser trigger signal. The energy of the laser pulse, $E_{Laser}$  in $\text{keV}_{\text{ee}}$, is deduced from the injected pulse length using the calibration obtained by comparing the laser NTD and ionization signal to those of the L and K lines.
The  SSED resistive response $R_{SSED}$ (in k$\Omega$) to various photon energies $E_{Laser}$ was measured as a function of two parameters: the applied bias $|\Delta V|$ which amplifies $E_{phonon}$ through NTL effect and the temperature difference $\Delta T = T_c - T_{op}$ which impacts the ability of an event to trigger the SSED.

\paragraph{SSED response as a function of the bias}

\begin{figure}[h!]
    \centering
    \includegraphics[width=0.49\linewidth]{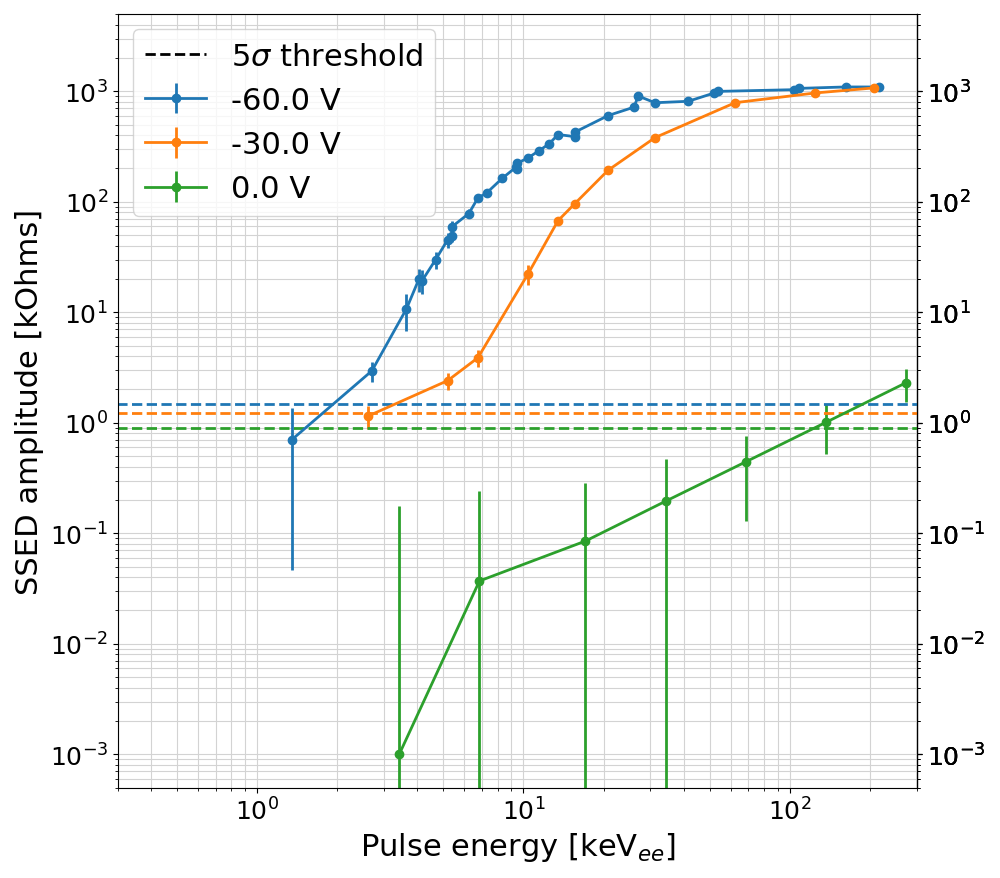}
    \includegraphics[width=0.49\linewidth]{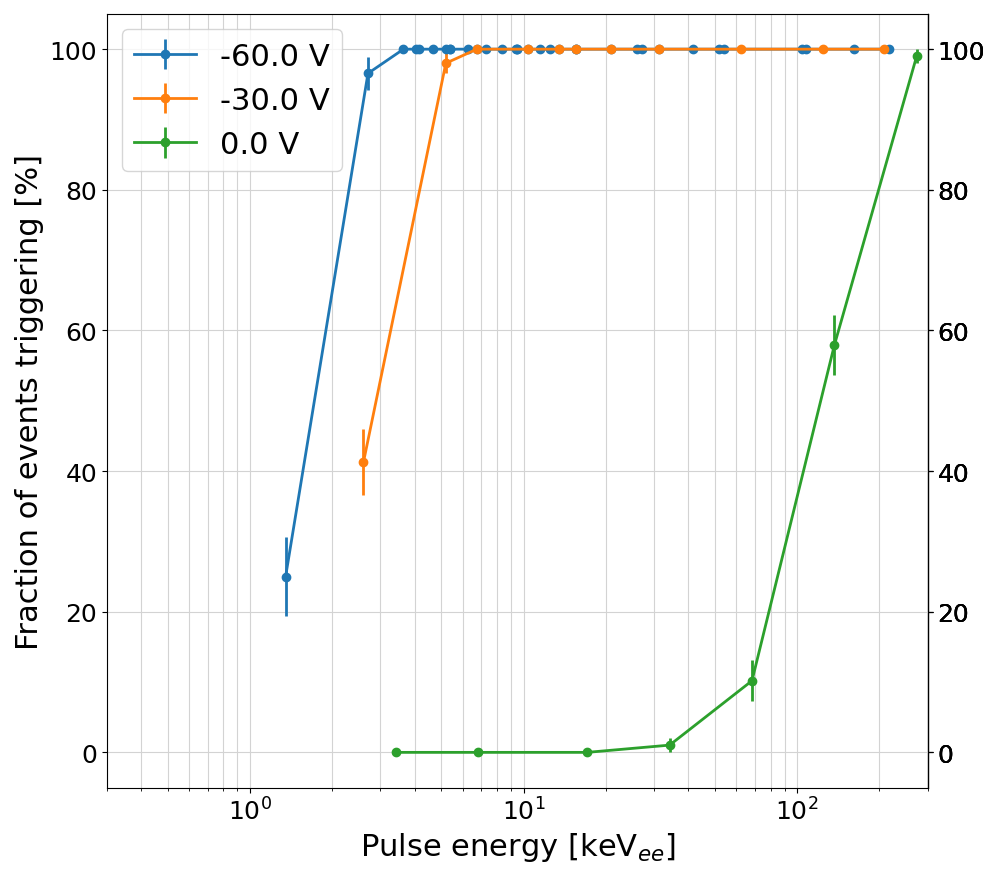}
    \caption{Left: SSED amplitude in k$\Omega$ as a function of the laser pulse energy in $\text{keV}_{\text{ee}}$ with $|\Delta V|=0$~V in green, $30$~V in orange and $60$~V in blue. The SSED amplitudes are extracted from the mean value of the Gaussian fit of the amplitude distributions per laser energy. The vertical error bars correspond to the statistical uncertainty of the fit values. The  dashed lines indicate the SSED threshold at 5$\sigma_{SSED} = 0.894~\text{k}\Omega$ (green), $1.215~\text{k}\Omega$ (orange), $1.489~\text{k}\Omega$ (blue), where $\sigma_{SSED}$ is the SSED baseline resolution (RMS). Right: Fraction of SSED events with $R_{SSED} > 5~\sigma_{SSED}$ as a function of the laser pulse energy in keV$_{ee}$. The vertical error bars correspond to a statistical uncertainty.}
    \label{fig-ssedbias}
\end{figure}

To study the SSED response to a varying bias, the CRYO50 detector is operated at a fixed temperature $T_{op} = 15$ mK which is chosen to be well below the critical temperature of the NbSi.
Laser pulses of various energies are injected in the crystal and Fig.~\ref{fig-ssedbias} (left) shows the amplitude of the pulse on the SSED in k$\Omega$ as a function of $E_{Laser}$ in $\text{keV}_{\text{ee}}$ for $|\Delta V| =$ 0~V (green), 30~V (orange) and 60~V (blue). 

It can be seen that the behavior of the SSED greatly depends on whether a bias is applied or not. For $|\Delta V| = 0$~V (green curve), the $e^- h^+$ pairs produced should not drift through the crystal, rather, they should recombine locally and, thus, should not trigger the SSED. The non-zero SSED amplitude at $|\Delta V|= 0$~V  at high energy are not observed in signals from gamma-ray interaction in the detector bulk. This behavior for laser pulse events can be explained by the fact that for each event, a fixed fraction of phonons is created in the crystal volume in the immediate vicinity of the SSED film. This interpretation is confirmed by the systematic presence at 0~V of a non-zero signal on the electrode for laser pulses ($\sim 0.5\%$ of the total charge), while it is not the case for bulk gamma-ray events. The few phonons reaching the film will affect its resistivity, but the signal remains below the threshold (defined as 5 times the baseline resolution on the SSED pulse amplitude) for $E_{Laser}<150$~ $\text{keV}_{\text{ee}}$.

When applying a bias, the dependence of the SSED signal on $E_{Laser}$ can be divided into three parts : the \textit{gap}, below which $E_{Laser}$ is not sufficient to produce enough NTL phonons to trigger the SSED above $5~\sigma_{SSED}$ (dashed lines on Fig~\ref{fig-ssedbias}, left). The second part is the \textit{threshold mode} or TES mode~\cite{refTES} for which $R_{SSED}$ is roughly proportional  to $E_{Laser}$. Lastly, the third regime is called \textit{saturation mode} where the entire SSED line undergoes a transition to its normal state and the signal reaches the maximal value of 1 $\text{M} \Omega$.

Increasing the bias increases the number of NTL phonons per IR photons. The $E_{Laser}$ value required to bring the SSED line halfway to its full transition ($R_{SSED} \approx 500~\text{k}\Omega$) is $18~\text{keV}_{\text{ee}}$ at 60~V and $E_{Laser} = 41~\text{keV}_{\text{ee}}$ at 30~V.
This increase by nearly a factor two mirrors that of the number of NTL phonons produced between 30 and 60 V.
The SSED is meant to act as a HO veto. On Fig.~\ref{fig-ssedbias} (right), we present the fraction of events triggering the SSED (i.e. $R_{SSED} > 5~\sigma_{SSED}$) as a function of $E_{Laser}$. At $|\Delta V| = 60$~V, the triggering efficiency reaches 100\% at $\approx$ 3.6 $\text{keV}_{\text{ee}}$, i.e. $\approx $ 840 $e^- h^+$ pairs. That number becomes  $\approx $ 1575 $e^- h^+$ pairs for $|\Delta V| = 30$~V, again mirroring the factor two difference in the production of NTL phonons.
The SSED threshold for electron recoils can clearly be improved by increasing the maximum bias that can sustain the detector, but other design factors will also contribute.

Another important factor is $\Delta T$. This has to be carefully controlled because the $\Delta T$ is crucial to veto the HO population. In the next part of this analysis, we study the influence of $\Delta T$ on the SSED triggering threshold with IR photons.

\paragraph{SSED response to varying $T_{op}$}

\begin{figure}[h!]
    \centering
    \includegraphics[width=0.49\linewidth]{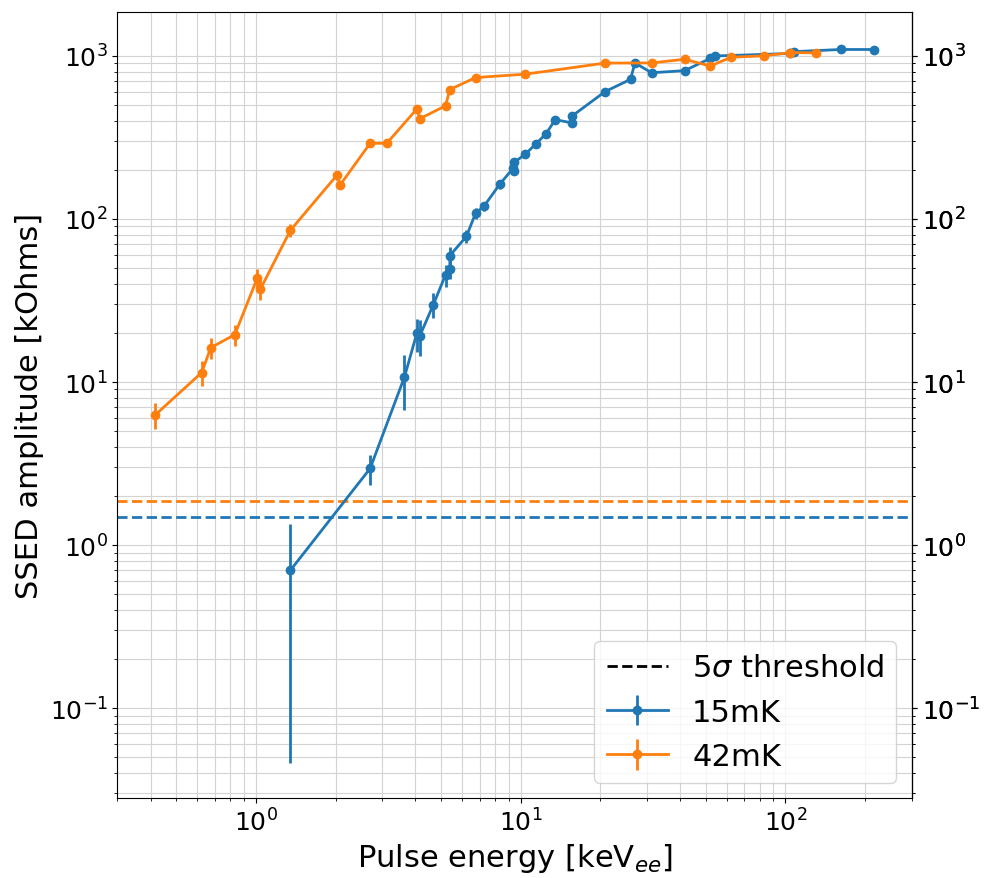}
    \includegraphics[width=0.49\linewidth]{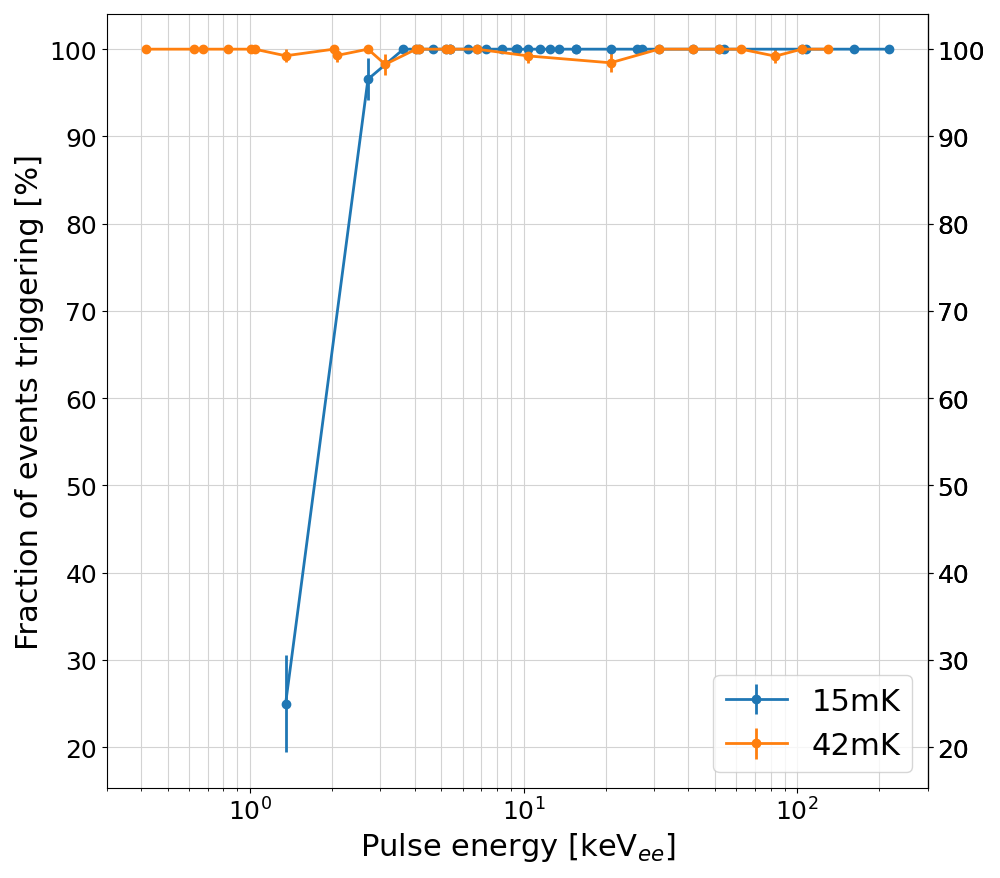}
    \caption{Data taken with $|\Delta V| = 60$~V. Left SSED amplitude in k$\Omega$ as a function of the laser pulse energy in $\text{keV}_{\text{ee}}$ with $T_{op} =15$~mK in blue and $42$~mK in orange. The SSED amplitudes are extracted from the mean value of the Gaussian fit of the amplitude distributions per laser energy. The vertical error bars correspond to the statistical uncertainty on the fit mean. The dashed lines indicate the triggering threshold at 5$\sigma_{SSED} = 1.489~\text{k}\Omega$ (blue) and $1.852~\text{k}\Omega$ (orange) for the two values of $T_{op}$, where $\sigma_{SSED}$ is the SSED baseline resolution (RMS). Right: Fraction of SSED events with $R_{SSED} > 5~\sigma_{SSED}$ as a function of the laser pulse energy in keV$_{ee}$. The vertical error bars correspond to the statistical uncertainty.}  
    \label{fig-ssedtemp}
\end{figure}

To study the SSED response to NTL phonon creation at various $T_{op}$, CRYO50 is operated at a fixed bias of $|\Delta V| = 60$~V.
Figure~\ref{fig-ssedtemp} (left) shows the SSED resistance as a function of the injected laser pulse energy for $T_{op} = 15$~mK in blue and $42$~mK in orange. The $5 \sigma_{SSED}$ threshold extracted from the SSED baseline resolution for both temperatures is represented with a dashed line of the corresponding color. 
Operating CRYO50 at a higher temperature lowers the SSED triggering threshold. 
The amount of energy required to bring the SSED line halfway to its full transition  is $E_{Laser} = 18~\text{keV}_{\text{ee}}$ at 15~mK and $E_{Laser} = 5~\text{keV}_{\text{ee}}$ at 42~mK.
A smaller temperature gap $\Delta T = T_c - T_{op}$ allows for smaller amounts of energy to trigger the SSED for a given $E_{Laser}$ energy. Indeed, at 15~mK, the SSED reach the $M\Omega$ range for higher pulse energies: around $E_{Laser} \approx 40~\text{keV}_{\text{ee}}$, against $E_{Laser} \approx 20~\text{keV}_{\text{ee}}$ at 42~mK. On Fig.~\ref{fig-ssedtemp} (right),  shows the corresponding $5~\sigma_{SSED}$ trigger efficiency. At $42$~mK, it is $100\%$ even for the lowest $E_{Laser}$ value probed (0.4 $\text{keV}_{\text{ee}}$). This highlights the importance of reducing the $T_C$ of the film to obtain the lowest possible threshold. This will be done in the next CRYOSEL prototype, now that the first one has established that the resolution on the NTD and SSED channel is sufficient to probe a lower energy domain.

\section{Conclusion}\label{sec3}

The response of the SSED sensor of the first CRYOSEL prototype, CRYO50, to IR laser have been studied in two scenarios. First, at a fixed temperature, varying the bias applied to the detector and thus yield of phonons created by the NTL effect. At 15~mK, the measured threshold for a nearly $100\%$ efficiency is $840$ $e^-h^+$ pairs at $60$~V. Secondly, the operating temperature has been varied at a fix bias of 60V. This  has shown that the difference of temperature between the operating temperature and the $T_C$ of the NbSi film is a key parameter to adjust to reach lower thresholds.

These first results are motivating to achieve the final objective of a sensitivity to a single $e^-h^+$ pair, they prove that the NbSi SSED can efficiently detect NTL phonons emitted in a small, very high-field region in front of it, thus identify ionizing events. Despite the very inhomogeneous field, the NTD channel achieves a very good peak energy resolution for the L and K lines of $^{71}$Ge at 1.3 and 10.37~keV, indicating a small dispersion of the phonon production as a function of the initial energy deposit. 
The path toward a single-electron sensitivity  goes by a lower $T_C$ of the SSED, an increase of the phonon collection efficiency by adapting the geometry of the SSED and Al electrodes, and finally an increase of the maximum bias that the detector can withstand without significant leakage current to directly increase the amount of NTL phonon.

\bmhead{Acknowledgments}
We acknowledge the support from the CRYOSEL collaboration, the EDELWEISS Collaboration, the RICOCHET Collaboration, Labex Lyon Institute of Origins (ANR-10-LABX-0066). The CRYOSEL project is supported in part by the French Agence Nationale pour la Recherche (ANR-21-CE31-0004).

\bibliography{sn-bibliography}

\end{document}